\documentclass[prl,twocolumn,floatfix]{revtex4-2}
\pdfoutput=1 
\usepackage{graphicx}
\usepackage{latexsym}
\usepackage{amsmath, amsthm, amssymb, wasysym}
\usepackage{xcolor}
\usepackage{epsfig}
\usepackage{soul}
\usepackage{cancel}
\usepackage{hyperref}
\usepackage{orcidlink}
\usepackage{pagecolor}
\usepackage{etoolbox}

\hypersetup{
 colorlinks=true,
 linkcolor=black,
 citecolor=blue,
 urlcolor=black
}

\definecolor{ForestGreen}{rgb}{0.10,0.45,0.20}
\definecolor{BurntOrange}{rgb}{0.75,0.35,0.10}

\begin{document}
%\setstcolor{red}
\title{  Boundary-Induced Drift and Negative Mobility in Constrained Stochastic Systems
}

\author{Meitar Goldfarb}

\author{Stanislav Burov\orcidlink{0000-0003-1065-174X}}

\email{stasbur@gmail.com}
\affiliation{Department of Physics, Bar-Ilan University, Ramat-Gan 5290002,
Israel}

\begin{abstract}
We study overdamped stochastic dynamics confined by hard reflecting boundaries and show that the combination of boundary geometry and an anisotropic diffusion tensor generically generates directed motion. 
At the level of individual trajectories, the no-flux condition enforces an oblique reflection at the boundary, which produces a systematic drift parallel to the surface. 
The resulting local velocity takes the general form
$v_B(\mathbf{x})=\mathbf{t}(\mathbf{x})^{\!\top}\mathbf{D}\,\mathbf{n}(\mathbf{x})$,
determined by the diffusion tensor $\mathbf{D}$ and the local boundary geometry encoded in the normal $\mathbf{n}$ and tangent $\mathbf{t}$. 
While this boundary-induced drift is local, it can accumulate into a macroscopic response, depending on the statistics of boundary encounters. 
We illustrate how this local boundary-induced drift gives rise to macroscopic transport using a minimal one-dimensional dimer composed of two particles with unequal diffusion coefficients.
The repeated collisions act as reflections in configuration space and lead to sustained center-of-mass motion, including regimes of absolute negative mobility under constant forcing.
\end{abstract}

\maketitle

\newcommand{\fr}{\frac}
\newcommand{\lr}{\langle}
\newcommand{\rl}{\rangle}
\newcommand{\tl}{\tilde}

%------------------ INTRODUCTION ------------------%

%Rectification of thermal fluctuations into directed motion is a central theme of nonequilibrium statistical physics. From Feynman’s ratchet to molecular motors, flashing potentials, and colloidal heat engines, microscopic systems are known to convert random thermal kicks into systematic transport by breaking detailed balance through spatial or temporal asymmetries~\cite{Reimann2002,Astumian2002,Hanggi2009,VanDenBroeck2004,Linke1999}. In such Brownian motors, symmetry breaking and energy input are inseparable: cyclic forcing, temperature gradients, or chemical reactions maintain steady probability currents~\cite{Filliger2007,Fogedby2017,Bechinger2016}. 
%The view that rectification necessarily requires an explicit nonequilibrium drive has become foundational in the design of synthetic microscale motors and active particles~\cite{Kay2007,Blickle2012,Tailleur2008,Hanggi2002}. 
%In equilibrium, detailed balance forbids any steady probability flux: spatial asymmetry may set a preferred direction, but directed motion requires a continuous input of energy or a violation of detailed balance.

Rectifying thermal fluctuations into directed motion typically requires an explicit nonequilibrium drive. 
Classical examples like ratchets driven by temporal modulation~\cite{Reimann2002}, 
molecular motors powered by chemical cycles~\cite{Astumian2002}, 
and colloidal engines operating across temperature gradients~\cite{Hanggi2009,VanDenBroeck2004,Linke1999} achieve transport by breaking detailed balance through external forcing.  
Related mechanisms also exploit cyclic forcing~\cite{Filliger2007}, 
controlled thermal gradients~\cite{Fogedby2017}, 
or active, chemically generated stresses in colloidal suspensions~\cite{Bechinger2016}.  
These principles underpin the design of synthetic microscale motors~\cite{Kay2007,Blickle2012} 
and chemically or mechanically driven active particles~\cite{Tailleur2008}. 
In equilibrium, detailed balance forbids steady currents: spatial asymmetry alone cannot produce directed motion.  
The appearance of constraint-induced  persistent motion in a fully passive system presents a qualitatively different form of rectification, which does not rely on temporal modulation or self-propulsion.
Such directed motion was observed by Bo and Eichhorn~\cite{BoEichhorn2017} for a Brownian particle in three dimensions near a reflecting wall subject to anisotropic yet homogeneous fluctuations.
Although the particle was driven only toward the wall and experienced no tangential forcing, a persistent drift parallel to the boundary emerged.  
This raises a general question: what physical principle allows anisotropic diffusion and hard reflection to generate persistent motion?

We show that in anisotropic diffusion processes, the zero–flux condition at a reflecting boundary necessarily induces an oblique reflection that generates a systematic drift along the surface.  
For a particle with a homogeneous diffusion tensor $\mathbf{D}$, and with $\mathbf{n}(\mathbf{x})$ the normal to the boundary at position $\mathbf{x}$, the drift in any direction $\mathbf{t}(\mathbf{x})$ parallel to the boundary is
%%%%%%%%%%%%%%%%%%%%%%%%%
\begin{equation}
v_{B}(\mathbf{x})
= \mathbf{t}(\mathbf{x})^{\!\top}\mathbf{D}\,\mathbf{n}(\mathbf{x}).
\label{eq:vBdefinition}
\end{equation}
%%%%%%%%%%%%%%%%%%%%%%%%%
This contribution is purely geometric, appears without any tangential forcing, and identifies the physical mechanism by which hard reflection and anisotropic diffusion can produce directed motion at the level of the effective stochastic dynamics.
To illustrate the implications of this tangential drift, we consider a minimal setting where the required anisotropy arises naturally: a one-dimensional dimer composed of two particles with different diffusion coefficients.
The boundary term in Eq.~\eqref{eq:vBdefinition} gives rise to regimes of persistent center-of-mass motion and negative mobility at long times. 

%We illustrate the implications of this boundary-induced drift in a setting where anisotropic diffusion naturally occurs, i.e., systems with particles with different self-diffusion coefficients. 
%Such systems show behavior, like the enhanced diffusion and directional memory, that do not appear for solutions of particles with identical diffusion coefficients.
%two minimal settings.  
%First, a single particle with anisotropic diffusion confined in a circular cavity and driven only by a radial force develops vortex-like steady currents generated solely by the geometric term.  
%Specifically, we show that a one-dimensional dimer composed of two particles with unequal diffusivities develops a directed center-of-mass motion, including regimes of negative mobility.  
%These examples demonstrate how the derived drift produces robust and counterintuitive transport effects in passive systems.

%Here, we answer this question by deriving the deterministic tangential velocity generated at any smooth reflecting boundary.  

%--------------------------------------------------%

\emph{Model.} We consider an overdamped Brownian particle confined to a domain 
$\Omega \subseteq \mathbb{R}^d$. 
In the interior, the dynamics is governed by the Itô stochastic differential equation
%%%%%%%%%%%%%%%%%%%%%%%%%%%%%
\begin{equation}
\mathrm{d}\mathbf{X}_t 
= \mathbf{b}(\mathbf{X}_t)\,\mathrm{d}t
+ \boldsymbol{\sigma}\,\mathrm{d}\mathbf{W}_t ,
\label{eq:Langevin_bulk}
\end{equation}
%%%%%%%%%%%%%%%%%%%%%%%%%%%%%%%
where $\mathbf{b}(\mathbf{x})$ is a drift field, 
$\boldsymbol{\sigma}$ is a homogeneous noise-amplitude matrix, 
and $\mathbf{W}_t$ is a vector of independent Wiener processes. 
The vectors are understood as column vectors, and below the inner products are written using the transpose notation.
As is well known, Eq.~\eqref{eq:Langevin_bulk} alone does not prevent trajectories from leaving $\Omega$. 
The positional probability density function $P(\mathbf{x},t)$ obeys the Fokker-Planck equation
%%%%%%%%%%%%%%%%%%%%%%%%%%%%%%
\begin{equation}
\partial_t P(\mathbf{x},t) 
= -\nabla\!\cdot (\mathbf{b}P(\mathbf{x},t))
+ \nabla\!\cdot\!\left(\mathbf{D}\nabla P(\mathbf{x},t)\right),
\label{eq:FP_bulk}
\end{equation}
%%%%%%%%%%%%%%%%%%%%%%%%%%%%%%%%%
where $\mathbf{D}=\tfrac12\,\boldsymbol{\sigma}\boldsymbol{\sigma}^{\!\top}$ is the diffusion tensor. Equation~\eqref{eq:FP_bulk} must be supplemented at $\mathbf{x}\in\partial\Omega$ by the no-flux (Neumann) boundary condition
%%%%%%%%%%%%%%%%%%%%%%%%%%%%%%
\begin{equation}
\mathbf{n\left(\mathbf{x}\right)}^{\!\top}\mathbf{J}=0,
\qquad 
\mathbf{J}=\mathbf{b}P(\mathbf{x},t)-\mathbf{D}\nabla P(\mathbf{x},t),
\label{eq:noflux}
\end{equation}
%%%%%%%%%%%%%%%%%%%%%%%%%%%
where $\mathbf{n\left(\mathbf{x}\right)}$ is the inward unit,  normal to the boundary $\partial\Omega$.
Eq.~\eqref{eq:noflux} ensures that the probability does not cross the boundary.  
The no-flux condition \eqref{eq:noflux} at the Fokker--Planck level must be 
accompanied by an equivalent constraint for individual trajectories.
This is implemented by the Langevin-Skorokhod construction
\cite{Skorokhod1961,Skorokhod1962,Tanaka1979,Lions1984},
%%%%%%%%%%%%%%%%%%%%%%%%%%
\begin{equation}
\mathrm{d}\mathbf{X}_t
= \mathbf{b}(\mathbf{X}_t)\,\mathrm{d}t
+ \boldsymbol{\sigma}\,\mathrm{d}\mathbf{W}_t
+ \mathbf{r}(\mathbf{X}_t)\,\mathrm{d}L_t .
\label{eq:Skorohod}
\end{equation}
%%%%%%%%%%%%%%%%%%%%%%%%%%
Equation~\eqref{eq:Skorohod} enforces the no-flux condition at the trajectory level, without introducing hard-wall discretization artifacts~\cite{behringer2011hard}.
Here $L_t$ is a non-decreasing process, proportional to the boundary local time, that increases only when $\mathbf{X}_t$ touches $\partial\Omega$~\cite{Tanaka1979,Lions1984}.  
The term $\mathbf{r}(\mathbf{x})\,\mathrm{d}L_t$ provides the minimal inward correction needed to keep the trajectory inside $\Omega$, by compensating the outward normal excursions generated by the Brownian increment $\boldsymbol{\sigma}\,\mathrm{d}\mathbf{W}_t$.  
Consistency with the no-flux condition fixes the reflection direction as~\cite{Lions1984,Saisho1987}
%%%%%%%%%%%%%%%%%%%%%%
\begin{equation}
\mathbf{r}(\mathbf{x})
=\frac{\mathbf{D}\,\mathbf{n}(\mathbf{x})}
       {\mathbf{n}(\mathbf{x})^{\!\top}\mathbf{D}\,\mathbf{n}(\mathbf{x})},
\quad \mathbf{x}\in\partial\Omega,
\label{eq:r_conormal}
\end{equation}
%%%%%%%%%%%%%%%%%%%%%%%
Equation~\eqref{eq:r_conormal} specifies the reflection direction at the boundary. 
For isotropic diffusion $\mathbf{D}\propto\mathbf{I}$, the reflection is normal to the surface, whereas anisotropy makes it oblique.

%%%%%%%%%%%%%%%%%%%%%%%%
%%%%Computation of v_{||}
%%%%%%%%%%%%%%%%%%%%%%%%%%%%

To compute the drift generated by the reflecting boundary, we analyze the short-time increment of the process \eqref{eq:Skorohod} and project it onto a direction parallel to the boundary. 
This corresponds to evaluating the local average of the tangential component of the displacement, which can be viewed as constructing a histogram of infinitesimal increments conditioned on $\mathbf{X}_t$ being located at the boundary point $\mathbf{x}$.
Integrating Eq.~\eqref{eq:Skorohod} over a short interval $\Delta t$ and using the Itô convention yields, to  leading order in $\Delta t$, that  
$\Delta\mathbf{X}_t=\mathbf{X}_{t+\Delta t}-\mathbf{X}_t
= \mathbf{b}(\mathbf{X}_t)\,\Delta t
+ \mathbf{\sigma}\,\Delta \mathbf{W}_t
+ \mathbf{r}(\mathbf{X}_t)\,\Delta L_t$, where $\Delta\mathbf{W}_t=\mathbf{W}_{t+\Delta t}-\mathbf{W}_t$ and $\Delta L_t=L_{t+\Delta t}-L_t$.
Let $\mathbf{t}(\mathbf{x})$ be a unit tangent vector to  $\partial \Omega$ at point $\mathbf{x}\in \partial \Omega$. 
We project the increment onto this direction, $\mathbf{t}(\mathbf{x}) ^{\!\top}\Delta \mathbf{X}_t/\Delta t$ and average over realizations of $\mathbf{X}_t$. 
The average $\mathbb{E}\left[\Delta\mathbf{W}_t\right]=0$ and to evaluate $\mathbb{E}\left[\mathbf{r}\left(\mathbf{X}_t\right)\Delta L_t\big/\Delta t\right]$ we use the occupation-layer representation of the Skorokhod term $L_t$~\cite{Tanaka1979} 
%%%%%%%%%%%%%%%%%%%%%
\begin{equation}
L_t = \lim_{\epsilon\to 0}\frac{1}{\epsilon} \int_0^t I\left(0<\phi(\mathbf{X}_s)<\epsilon\right) \mathbf{n}(\mathbf{X}_s)^{\!\top} \mathbf{D}\mathbf{n}(\mathbf{X}_s)ds
    \label{eq:Ltdefinition}
\end{equation}
%%%%%%%%%%%%%%%%%%%%
%$L_t = \lim_{\epsilon\to 0}\frac{1}{\epsilon} \int_0^t \mathbf{1}_{0<\phi(\mathbf{X}_s)<\epsilon} \mathbf{n}(\mathbf{X}_s)^{\!\top} \mathbf{D}\mathbf{n}(\mathbf{X}_s)ds$, 
where $I(...)$ is the indicator function and $\phi(\mathbf{X}_s)$ is the minimal distance between $\partial \Omega$ and $\mathbf{X}_s$.
%$\mathbf{1}_{A}$ is $1$ when $A$ is true and $0$ otherwise.   $\phi(\mathbf{X}_s)$ is the minimal distance between $\mathbf{X}_s$ and the boundary $\partial \Omega$. 
As $\Delta t\to 0$,  the increment satisfies $\Delta L_t\big/{\Delta t}\to \lim_{\epsilon\to 0}\frac{1}{\epsilon}I\left(0<\phi(\mathbf{X}_t)<\epsilon\right) \mathbf{n}(\mathbf{X}_t)^{\!\top}\mathbf{D}\mathbf{n}(\mathbf{X}_t)$ and the  Skorokhod term 
$\mathbb{E}\left[\mathbf{t}(\mathbf{x})^{\!\top}\mathbf{r}\left(\mathbf{X}_t\right)\Delta L_t\big/\Delta t\right]$ converges to $\int_{\Omega} \lim_{\epsilon\to 0}\epsilon^{-1}I\left(0<\phi(\mathbf{\tilde x})<\epsilon\right)  \mathbf{t}(\mathbf{x})^{\!\top}\mathbf{D}\mathbf{n}\left(\mathbf{\tilde x}\right)P(\mathbf{\tilde x},t)\,d\mathbf{\tilde x}$, where we used the form in Eq.~\eqref{eq:r_conormal} for $\mathbf{r}(\mathbf{X}_t)$.
%and $P(\mathbf{\tilde x},t)$ is the  probability density function (PDF) at point $\mathbf{\tilde x}$. 
Because $\lim_{\epsilon\to 0}\epsilon^{-1}I\left(0<\phi(\mathbf{\tilde x})<\epsilon\right)\to \delta(\phi(\mathbf{\tilde x}))$ the volume integral collapses to the boundary and the  drift $\mathbf{v}(t)=\lim_{\Delta t\to0} \mathbb{E}\left[\Delta \mathbf{X}_t/\Delta t\right]$ projected on $\mathbf{t}(\mathbf{x})$ is
%%%%%%%%%%%%%%%%%%%%%%
\begin{equation}
\begin{array}{l}
   \mathbf{t}(\mathbf{x})^{\!\top}\mathbf{v}(t) = 
    \\
       \int_{\Omega}  \mathbf{t}(\mathbf{x})^{\!\top}b(\mathbf{\tilde x})P\left(\mathbf{\tilde x},t\right)\,d\mathbf{\tilde x} + 
    \int_{\partial \Omega}  \mathbf{t}(\mathbf{x})^{\!\top}\mathbf{D}\mathbf{n}(\mathbf{\tilde x})P_B(\mathbf{\tilde x},t)\,dS_{\mathbf{\tilde x}}
    \end{array}
    \label{eq:overalAverage}
\end{equation}
where $P_B(\mathbf{\tilde x},t)$ is the PDF at the boundary and $S_{\mathbf{\tilde x}}$ is the surface element of the boundary. 
Conditioning on $\mathbf{X}_t=\mathbf{x}\in\partial\Omega$, and applying the law of total expectation, Eq.~\eqref{eq:overalAverage} yields for the local drift along $\mathbf{t}(\mathbf{x})$ 
%For the average drift at point $\mathbf{x}\in\partial \Omega$ we use the conditional expectation $v_{\mathbf{t}(\mathbf{x})} = \mathbb{E}\left[\mathbf{t}(\mathbf{x})^{\!\top}\mathbf{v}(\mathbf{X}_t)|\mathbf{X}_t=x\right]$ and according to the law of total expectation, Eq.~\eqref{eq:overalAverage} yields
%%%%%%%%%%%%%%%%%%%%
\begin{equation}
v_{\mathbf{t}(\mathbf{x})} = \mathbf{t}(\mathbf{x})^{\!\top} \mathbf{b}(\mathbf{x}) + \mathbf{t}(\mathbf{x})^{\!\top}\mathbf{D}\mathbf{n}(\mathbf{x})
    \label{eq:vparalllelGen}
\end{equation}
%%%%%%%%%%%%%%%%%%%%
Equation~\eqref{eq:vparalllelGen} holds for arbitrary smooth reflecting boundaries and homogeneous diffusion tensors, independent of the global geometry of the confining domain.
The second term on the right-hand side of Eq.~\eqref{eq:vparalllelGen} is $v_B$ in Eq.~\eqref{eq:vBdefinition}. This additional term $v_B$ is the local velocity injected by the wall, an unavoidable drift produced whenever anisotropic noise meets a hard boundary.
For isotropic diffusion, $v_B$ disappears, but in the anisotropic case, it can become the source of nonconventional transport effects. 

%%%%%%%%%%%%%%%%%%%%%%%

\emph{Dimer with unequal diffusivities.}
The boundary-induced drift $v_B$ derived above is a local effect that acts only when the particle touches the reflecting surface.  
To generate a sustained macroscopic current, these reflection events must occur repeatedly.  
A single particle with anisotropic diffusion next to a wall makes only sporadic contacts and cannot accumulate a net drift.  
A dimer composed of two particles held in close proximity by an external compressive force provides the minimal setting for repeated reflection events, i.e., each collision between the two particles acts as a reflection at a moving boundary.  
When particles have unequal diffusion coefficients, the dimer acquires an effective anisotropy, making it a simple and experimentally accessible realization of the present setting~\cite{berut2014energy,ciliberto2025experimental}.
%For example, in colloidal pairs confined in narrow channels or on optical-tweezer rings[CITE].

We model the dimer as two overdamped Brownian particles with coordinates  $x_A(t)$ and $x_B(t)$, with particle $A$ constrained to remain to the right of particle $B$.  
They are driven toward each other by constant generalized forces $F_A$ and $F_B$, and diffuse with coefficients $D_A$ and $D_B$, respectively.  
Their stochastic dynamics in the absence of interactions is  

%%%%%%%%%%%%%%%%%%%%%%%%%%%%%%%%%%%
\begin{equation}
\begin{array}{l}
\mathrm{d}x_A = F_A\,\mathrm{d}t + \sqrt{2D_A}\,\mathrm{d}W_A(t)
\\
\mathrm{d}x_B = F_B\,\mathrm{d}t + \sqrt{2D_B}\,\mathrm{d}W_B(t)
\end{array}
\label{eq:dimer_langevin}
\end{equation}
%%%%%%%%%%%%%%%%%%%%%%%%%%%%%%%%%%%%
where $W_A$ and $W_B$ are independent Wiener processes.  
The hard–core interaction imposes the constraint  
$x_A(t) > x_B(t)$,
forbidding overtaking and producing point-like collisions between the two particles. 
The joint probability density $P(x_A,x_B,t)$ obeys the Fokker-Planck equation
%%%%%%%%%%%%%%%%%%%%%%%%%
\begin{equation}
\partial_t P
= 
-\partial_{x_A}(F_A P) - \partial_{x_B}(F_B P)
  + D_A\,\partial_{x_A}^2 P + D_B\,\partial_{x_B}^2 P
\label{eq:dimer_FP}
\end{equation}
%%%%%%%%%%%%%%%%%%%%%%%%%%%%%%%
defined on the half-plane
$
\Omega = \{x_A>x_B\}$ with the reflecting boundary  $\partial\Omega =\{x_A=x_B\}$.
The associated probability current is $\mathbf{J}=(J_A,J_B)^{\!\top}$ while
$
J_A = F_A P - D_A\,\partial_{x_A}P
$ and
$J_B = F_B P - D_B\,\partial_{x_B}P$.
The dynamics satisfy the no-flux condition $\mathbf{n}\cdot\mathbf{J}=0$ at $\partial \Omega$, 
where $\mathbf{n}$ is the inward unit normal to $\partial\Omega$, i.e., $\mathbf{n}=(1,-1)^{\!\top}/\sqrt{2}$.

Equation~\eqref{eq:dimer_FP} describes a diffusion process in the $(x_A,x_B)$ plane with a homogeneous diffusion tensor
%%%%%%%%%%%%%%%%%%%%%%%%%
\begin{equation}
\mathbf{D}_{\mathrm{dimer}} =
\begin{pmatrix}
D_A & 0 \\
0   & D_B
\end{pmatrix},
\label{eq:dimer_Dtensor}
\end{equation}
%%%%%%%%%%%%%%%%%%%%%%%%%%%
confined by a reflecting boundary along $x_A = x_B$.  
When $D_A \neq D_B$, the tensor $\mathbf{D}_{\mathrm{dimer}}$ is anisotropic, and the collision line plays the role of the reflecting boundary in the general mechanism discussed above.  
In this representation, the pair $(x_A,x_B)$ diffuses in the $(x_A,x_B)$ plane with a hard reflecting boundary and anisotropic diffusion. Therefore, the boundary-induced drift term 
$v_B = \mathbf{t}^{\!\top}\mathbf{D}\mathbf{n}$ from Eq.~\eqref{eq:vBdefinition} applies locally to the dimer dynamics.

The direction along the boundary is $\mathbf{t}=(1,1)^{\!\top}/\sqrt{2}$. 
Substituiting $\mathbf{t}$, $\mathbf{n}$, $\mathbf{b}=(F_A,F_B)^{\!\top}$ and the diffusion tensor $\mathbf{D}_{\mathrm{dimer}}$ into Eq.~\eqref{eq:vparalllelGen} gives the drift along the boundary  $v_{||} = \left(F_A+F_B\right)/\sqrt{2}+\left(D_A-D_B\right)/2$. 
The first term is the projection of the external generalized forces along the boundary, while the second term, $v_B=\left(D_A-D_B\right)/2$, is the geometric boundary-induced contribution arising from the oblique reflection generated by the anisotropic diffusion tensor. 
On the level of two single particles in the dimer, $v_B\neq 0$  reflects the fact that collisions do not reverse the inter-particle motion symmetrically. This asymmetry disappears only when $D_A=D_B$, for which the reflection becomes normal and the geometric drift vanishes. 

%To generate a macroscopic current, the boundary-induced drift $v_{B}$ must be sampled repeatedly.
%From Eq.~\eqref{eq:overalAverage}, its global contribution is provided by 
%$\int_{\partial\Omega}\mathbf{t}^{\!\top}\mathbf{D}_{\mathrm{dimer}}\mathbf{n}\,P_{B}(\tilde{x},t)\,dS_{\tilde{x}}$ 
%so a persistent effect requires that the probability of being on the boundary, i.e, $c(t)=\int_{\partial\Omega}P_{B}(\tilde{x},t)\,dS_{\tilde{x}},$, 

\begin{figure}[t]
 \centering
 \includegraphics[width=0.99\linewidth]{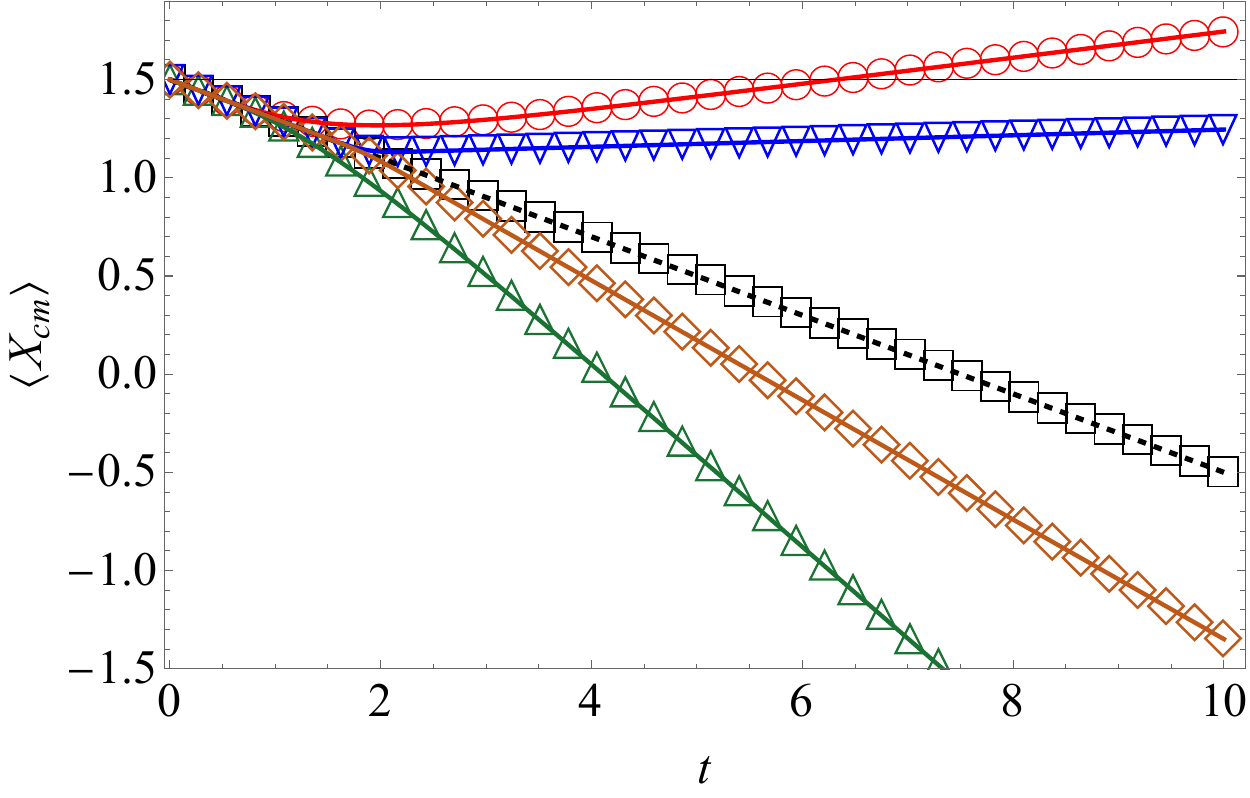}  
 \caption{
 Time evolution of the dimer center-of-mass position $X_{\mathrm{cm}}(t)$ under a fixed negative total generalized force $F_A+F_B<0$.
For certain diffusion-coefficient ratios, the center of mass moves against the applied force, $X_{\mathrm{cm}}(t)$ increasing with time, demonstrating absolute negative mobility ($\textcolor{red}\bigcirc$, $\textcolor{blue}\bigtriangledown
$).
Other choices of $D_A$ and $D_B$ lead to conventional drift in the force direction ($\Box$,$\textcolor{BurntOrange}\Diamond$,$\textcolor{ForestGreen}\bigtriangleup$). 
The horizontal black line marks the initial center-of-mass position and highlights motion against the applied force.
Symbols show numerical simulations, while thick solid lines are analytical predictions obtained from Eq.\eqref{eq:vcmAveragegl} (with $c(t)$ given in Eq.(A14) of the SM). Specifically $\textcolor{red}\bigcirc$: ($D_A=0.5$, $D_B=0.25$),  $\textcolor{blue}\bigtriangledown
$: ($D_A=0.013$, $D_B=0.0075$), $\Box$: ($D_A=D_B=0.5$), $\textcolor{BurntOrange}\Diamond$: ($D_A=0.01$, $D_B=0.013$), and  $\textcolor{ForestGreen}\bigtriangleup$: ($D_A=0.25$, $D_B=0.5$). 
For all cases $F_A=-1$ and $F_B=0.6$. 
The dashed line indicates the short-time behavior, which persists only in the isotropic case where $D_A=D_B$ ($\Box$).
Results of numerical simulations were averaged over $10^5$  realizations and $\Delta t = 10^{-3}$ (additional details are in SM).
 %Despite identical forcing ($F_A=-1$ and $F_B=0.6$), the system response is sensitive to the diffusion coefficients $D_A$ and $D_B$. For certain diffusivity ratios ($\textcolor{red}\bigcirc$ - $D_A=0.5$, $D_B=0.25$ and $\textcolor{blue}\bigtriangledown$ - $D_A=0.013$, $D_B=0.0075$), $X_{cm}$ increases with time, demonstrating absolute negative mobility, while other choices lead to conventional drift ($\Box$ - $D_A=D_B=0.5$; $\textcolor{BurntOrange}\Diamond$ - $D_A=0.01$, $D_B=0.013$; $\textcolor{ForestGreen}\bigtriangleup$ - $D_A=0.25$, $D_B=0.5$;).
%The horizontal black line indicates the initial center-of-mass position, providing a reference for motion against the applied force.
%Symbols represent numerical simulations for different $D_A$ and $D_B$, while thick solid lines are analytical predictions obtained from integration of Eq.~\eqref{eq:vcmAveragegl} ($c(t)$ is provided in Eq. (A14) of the SM). 
%The dashed line shows the short-time behavior, which remains uninterrupted only in the isotropic case $D_A=D_B$ ($\Box$).
%Results of numerical simulations were averaged over $10^5$  realizations and $\Delta t = 10^{-3}$ (additional details are in SM).
}\label{fig:anmfig}
\end{figure}

The existence of a non-zero $v_B$ doesn't, in itself, guarantee the emergence of macroscopic flows. 
It states that along the boundary, the geometrical contribution $v_B$ can compete with the projected deterministic drift.  Indeed, from Eq.~\eqref{eq:overalAverage}, we see that the global impact of $v_B$ depends on the term $\int_{\partial \Omega} \mathbf{t}^{\!\top}\mathbf{D}_{\mathrm{dimer}}\mathbf{n} P_{B}(\tilde{x},t)\,dS_{\tilde{x}}$, i.e., the particle has to spend signifcant amount of time in the vicinity of the boundary $\partial \Omega$ to compete with the regular drift due to $\mathbf{b}$. 
For the dimer, the average motion of the center of mass $X_{cm}=\left(x_A+x_B\right)/2$ is exactly an example of where the presence of non-zero $v_B$ can lead to macroscopic effects. 
Since $(x_A,x_B)\mathbf{t} =\sqrt{2} X_{cm}$, and $\mathbf{t}$, $\mathbf{n}$, and $\mathbf{D}_{\mathrm{dimer}}$ are constant, Eq.~\eqref{eq:overalAverage} yields
%%%%%%%%%%%%%%%%%%%%%%%%%%%%%%%%%%%%%%
\begin{equation}
\langle \dot{X}_{cm}\rangle = \frac{F_A+F_B}{2} + \frac{D_A-D_B}{2\sqrt{2}} c(t)
    \label{eq:vcmAveragegl}
\end{equation}
%%%%%%%%%%%%%%%%%%%%%%%%%%%%%%%%%%%%%%
where $c(t)=\int_{\partial \Omega} P_B(\tilde{x},t)\,dS_{\tilde{x}}$ is the probability mass of presence at the boundary. 
The term on the right-hand side determines how strong the effect of differences in diffusivities will be, if at all. 
The resulting time evolution of $X_{cm}$, including regimes where the motion occurs against the applied force, is illustrated in Fig~\ref{fig:anmfig}.
The two knobs of the effect are the difference $D_A-D_B$ and $c(t)$. To observe a long-lasting effect, $c(t)$ shouldn't converge to $0$ as a function of time. Namely, the generalized forces $F_A$ and $F_B$ should act such that in the $(x_A,x_B)$ plane the particle is constantly present in the vicinity of the boundary $x_A=x_B$, i.e., the collisions of the two particles should occur frequently and not as rare events. 
In the Supplemntal Material we compute $c(t)$ for arbitrary $t,D_A,D_B,F_A,F_B$, by finding the PDF of the relative coordinate $x_A-x_B$.
In the limit of short times $t\to 0$, the effect of $v_B$ is non existent and $\langle \dot{X}_{cm}\rangle =\frac{1}{2} (F_A+F_B)t$, as long as the initial state is not $x_A=x_B$. 
In the limit of long times, $t\to \infty$, $c(t)$ is non-zero only when $(F_A,F_B)\cdot\mathbf{n}<0$, i.e. $F_A-F_B<0$, the projection of the generalized force on the direction perpendicular to the boundary $x_A=x_B$ is negative. 
As we stated above, the particles have to be driven towards each other. Under this condition, we obtain
%%%%%%%%%%%%%%%%%%%%%%%%%%%%%%%%%%%%%%%
\begin{equation}
    \langle X_{cm}\rangle \sim  \left(\frac{F_A+F_B}{2}  + \frac{D_A-D_B}{2}\frac{F_B-F_A}{D_A+D_B}
    \right)t
    \label{eq:xcmPosgen}
\end{equation}
%%%%%%%%%%%%%%%%%%%%%%%%%%%%%%%%%%%%%%
while $F_A$ and $F_B$ can attain positive or negative values.
In this regime, the average distance between the particles converges to a constant (see SM)  %$\langle x_A-x_B \rangle =\left(D_A+D_B\right)\big/|\sqrt{2}(F_A-F_B)|$, 
indicating that the particles remain in frequent contact rather than separating for extended periods of time.

%%%%%%%%%%%%%%%%%%%%%%%%%%%%%
%%%%%%%%%%%%%%%%%%%%%%%%%%%%%%%%%%%%%%%%%%%

Equation~\eqref{eq:xcmPosgen} displays that a non-zero $v_B$ gives rise to a situation where the total external generalized force $F_A+F_B=0$, but there is still a systematic motion of the center of mass. This occurs whenever $F_B-F_A>0$. The direction of the motion is defined by the sign of $D_A-D_B$, i.e., from the particle with the lower diffusion coefficient and towards the particle with the higher diffusion coefficient. 
Moreover, even when $F_A+F_B\neq 0$, the behavior of $\langle X_{cm}\rangle$ provides unexpected features. 
The expected direction of motion is defined by the sign of $F_A+F_B$, but the additional term in Eq.~\eqref{eq:xcmPosgen} can flip the direction of motion. 
Specifically, when $|F_A+F_B|<|(D_A-D_B)(F_B-F_A)|/(D_A+D_B)$ the phenomenon of absolute negative mobility (ANM)~\cite{reimann1999coupled,eichhorn2002brownian,Reimann2002,ros2005absolute} can occur. 
ANM describes situations in which the system's response is opposite to the direction of the applied external force. 
The appearance of ANM is often associated with time-dependent external forcing~\cite{du2012absolute,dandogbessi2015absolute}. 
But in Eq.~\eqref{eq:xcmPosgen}, ANM arises from asymmetric reflection events induced by diffusional anisotropy in the $(x_A,x_B)$ space at the level of the effective stochastic dynamics.
Our predictions are in excellent agreement with numerical simulations, as shown in Fig.~\ref{fig:anmfig}. 

Up to this point, the dimer example has been analyzed with generalized forces and diffusivities as independent parameters of an effective stochastic description. 
This level of generality is appropriate for identifying the geometric mechanism underlying boundary-induced drift. 
However, additional constraints arise when the stochastic dynamics is required to be compatible with equilibrium statistical mechanics.
For particles coupled to thermal reservoirs characterized by temperatures $T_i$ and friction coefficients $\gamma_i$~\cite{grosberg2015nonequilibrium,grosberg2018dissipation,PhysRevE.106.064608}, the fluctuation--dissipation relation imposes
$D_i=k_BT_i/\gamma_i$ and $F_i=f_i/\gamma_i$, where $f_i$ is the physical force acting on particle $i$. 
Substituting these relations into Eq.~\eqref{eq:xcmPosgen}, the center-of-mass velocity attains the asymptotic form
%%%%%%%%%%%%%%%%%%%%%%%%%%%%
\begin{equation}
\langle  X_{cm} \rangle
\sim
\frac{T_A f_B + T_B f_A}{T_A\gamma_B + T_B\gamma_A} t.
\label{eq:Xcm_twoT}
\end{equation}
%%%%%%%%%%%%%%%%%%%%%%%%%%%%%%
Equation~\eqref{eq:Xcm_twoT} shows that spontaneous currents and ANM are generically allowed when the two particles are coupled to distinct thermal environments.
In contrast, when both particles are coupled to the same thermal reservoir, $T_A=T_B\equiv T$, the asymptotic center-of-mass velocity reduces to
$(f_A+f_B)/(\gamma_A+\gamma_B)$ and vanishes whenever the total physical force is zero.
Thus, while anisotropic reflection universally generates a local boundary-induced drift, thermal equilibrium emerges as the unique situation in which its macroscopic effect on the center-of-mass motion cancels in the long-time limit.

\emph{Discussion.} The results presented here establish a general geometric mechanism by which constrained stochastic dynamics can generate systematic transport. 
Anisotropic diffusion combined with hard reflection injects a local tangential drift at boundaries through oblique reflection, independent of temporal modulation, memory, or self-propulsion, as quantified by Eq.~\eqref{eq:vBdefinition}. 
While this boundary-induced drift is inherently local, its macroscopic impact is controlled by the statistics of boundary encounters and can therefore persist at long times in appropriately constrained systems.

The dimer provides a minimal realization of this principle, where repeated collisions act as reflections in configuration space and allow the accumulation of the boundary-induced drift into a sustained center-of-mass response, including regimes of absolute negative mobility. 
This mechanism is distinct from transport phenomena in many-body mixtures with heterogeneous constituents~\cite{tanaka2017hot,ilker2021long}, in which directed motion has been shown to arise from environment-mediated interactions and collective effects~\cite{schwarcz2024emergence,benois2023enhanced,al2025cold}. 
Here, transport emerges solely from the geometry of the constrained stochastic dynamics.
The appearance of absolute negative mobility, sustained by frequent collisions, has also been reported in driven robotic systems, where a vibrated robot interacting with an inclined plane climbs against gravity through repeated collisions with the boundary~\cite{ben2023morphological}.
We expect that the geometric rectification mechanism discussed here will give rise to unexplored transport phenomena in systems with strong quenched disorder~\cite{shafir2024disorder}.

Importantly, our results do not contradict equilibrium expectations. 
When fluctuation–dissipation relations enforce thermal equilibrium, the long-time contribution of the boundary-induced drift to the center-of-mass motion cancels, and no steady macroscopic current remains. 
Equilibrium thus appears as the unique limit in which this local mechanism leaves no global trace. 

\begin{acknowledgments}
{\emph{Acknowledgments-}} This work was supported by the Israel Science Foundation, Grant No.~3791/25.
\end{acknowledgments}

%More broadly, this work shows how hard constraints and anisotropic fluctuations can combine to generate systematic transport through purely geometric effects at boundaries.

%\emph{Discussion.}
%Beyond this minimal model, systems containing particles with unequal diffusivities exhibit additional nonequilibrium transport phenomena[CITE].  
%In many-body mixtures, repeated collisions with surrounding particles can transiently confine two dissimilar particles in close proximity, producing correlated pair motion[CITE] and enhanced diffusion[CITE].
%Although the phenomenology—directed or persistent motion of a heterogeneous pair—is similar, the underlying mechanisms differ: in mixtures the drift originates from environment-mediated collisions, whereas in the dimer studied here the collisions occur only between the two constituents and generate transport through the boundary-induced geometric drift identified in Eq.~\eqref{eq:vBdefinition}.
%This contrast highlights a broader principle: heterogeneous fluctuations, when combined with sustained proximity, can generically produce nontrivial transport, yet the geometric mechanism presented here operates even in fully passive, Markovian systems without many-body forcing.

%\section{another introduction}

%\section{Introduction}

%%%%%%%%%%%%%%%%%%%%%%%%%%%%%
%%%%%%%%%%%%%%%%%%%%%%%%%%%%%
%%%%%%%%%%%%%%%%%%%%%%%%%%%%%
%%%%%%%%%%%%%%%%%%%%%%
%Every boundary collision induces an average displacement equal to D(x)n(x)\,dt.

%Only the tangential part survives reflection, giving a mean drift t^T Dn.”**

%%%%%%%%%%%%%%%%%%%%%%

\bibliography{activitywithout.bib}

\end{document}

%\setstcolor{red}
\title{ Supplemental Material for: Boundary-Induced Drift and Negative Mobility in Constrained Stochastic Systems
}

\author{Meitar Goldfarb}

\author{Stanislav Burov\orcidlink{0000-0003-1065-174X}}

\email{stasbur@gmail.com}
\affiliation{Department of Physics, Bar-Ilan University, Ramat-Gan 5290002,
Israel}

\begin{abstract}
The Supplemental Material consists of two sections: In Sec. A., we derive an exact expression for the boundary occupation $c(t)$, which controls the emergence of macroscopic drift in the dimer model. In Sec. B. we specify the details of numerical simulations.
\end{abstract}

\maketitle

\newcommand{\fr}{\frac}
\newcommand{\lr}{\langle}
\newcommand{\rl}{\rangle}
\newcommand{\tl}{\tilde}

\section{ Section A.~ Derivation of $c(t)$ in the dimer model.}

\setcounter{equation}{0}
\renewcommand{\theequation}{A\arabic{equation}}

The Fokker-Planck equation for $P(x_A,x_B;t)$ is provided by Eq.~(11), namely 
%%%%%%%%%%%%%%%%%%%%%%%%%
\begin{equation}
\partial_t P
= 
-\partial_{x_A}(F_A P) - \partial_{x_B}(F_B P)
  + D_A\,\partial_{x_A}^2 P + D_B\,\partial_{x_B}^2 P
\label{eq:genFP}
\end{equation}
%%%%%%%%%%%%%%%%%%%%%%%%%%%%%%%
The zero flux condition at $x_A=x_B$ is provided by 
%%%%%%%%%%%%%%%%%%%%%%%%%%%%%%%
\begin{equation}
-F_A P+D_A \frac{\partial P}{\partial {x_A}}+F_B P-D_B\frac{\partial P}{\partial x_B}
=0, 
    \label{eq:zerofluxGen}
\end{equation}
%%%%%%%%%%%%%%%%%%%%%%%%%%%%%%%
at $x_A=x_B$. First, we derive the solution of the Fokker-Planck equation in Fourier-Laplace space. 

We assume that at $t=0$, $P(x_A,x_B,t=0)=\delta(x_A-x_A^O)\delta(x_B-x_B^O)$. 
We transform to coordinates  $u=\left(x_A+x_B\right)/\sqrt{2}$ and $w=\left(x_A-x_B\right)/\sqrt{2}$, while $F_u =\left(F_A+F_B\right)/\sqrt{2}$ and $F_w=\left(F_A-F_B\right)/\sqrt{2}$. 
In the following, we assume that the particles in the dimer are pushed towards each other, i.e., $F_w<0$. 
In these new coordinates for $w\geq 0$ Eq.~\eqref{eq:zerofluxGen} attains the form 
%%%%%%%%%%%%%%%%%%%%%%%%%%%%%%%%%%%%%%%%%%%%%% 
\begin{equation}
\begin{array}{l}
\frac{\partial P}{\partial t} = 
-F_w\frac{\partial P}{\partial w} 
+
 \frac{1}{2}(D_A+D_B) \left( \frac{\partial^2 P}{\partial w^2}+\frac{\partial^2 P}{\partial u^2} \right)
-
F_u\frac{\partial P}{\partial u} 
\\
+
(D_A-D_B) \frac{\partial^2 P }{\partial w \partial u},
\end{array}
    \label{eq:transformedfokkerplanck}
\end{equation}
%%%%%%%%%%%%%%%%%%%%%%%%%%%%%%%%%%%%%%%%%%%%%%%
while the initial condition is $P(u,w;t=0)=\delta(u-u^O)\delta(w-w^O)$.
Next, we apply the Laplace transform, i.e, $\int_0^\infty e^{-st}\left(\dots\right)\,dt$, to both sides of Eq.~\eqref{eq:transformedfokkerplanck} and obtain
%%%%%%%%%%%%%%%%%%%%%%%%%%%%%%%%%%%%%%%
\begin{equation}
\begin{array}{ll}
s\hat{P}-\delta(u-u^O)\delta(w-w^O) =
-F_w\frac{\partial \hat{P}}{\partial w} 
+
\\
\frac{1}{2}(D_A+D_B) \left( \frac{\partial^2 P}{\partial w^2}+\frac{\partial^2 \hat{P}}{\partial u^2} \right)
-
F_u\frac{\partial \hat{P}}{\partial u} 
+
(D_A-D_B) \frac{\partial^2 \hat{P} }{\partial w \partial u},
\end{array}
    \label{eq:fpLpalcefirst}
\end{equation}
%%%%%%%%%%%%%%%%%%%%%%%%%%%%%%%%%%%%%%%
where $\hat{P} = \int_0^\infty P(u,w;t)e^{-st}\,dt$. Subsequent application of Fourier transform to both sides, i.e., $\int_{-\infty}^\infty e^{-iku}\left(\dots\right)\,du$, leads to 
%%%%%%%%%%%%%%%%%%%%%%%%%%%%%%%%%%%%%%%%
\begin{equation}
\alpha \frac{\partial^2 \hat{\tilde{P}}}{\partial w^2}+\beta \frac{\partial \hat{\tilde{P}}}{\partial w}+\gamma\hat{\tilde{P}} = -\delta(w-w^O)e^{-iku^O}
    \label{eq:fpafterLFt}
\end{equation}
%%%%%%%%%%%%%%%%%%%%%%%%%%%%%%%%%%%%%%%%%
where $\tilde{P} = \int_{-\infty}^{\infty} P(u,w;t) e^{-iku}\,du$ and 
%%%%%%%%%%%%%%%%%%%%%%%%%%%%%%%%%%%%%%
\begin{equation}
\begin{array}{l}
    \alpha = \frac{1}{2}\left(D_A+D_B\right)
    \\
    \beta = ik\left(D_A-D_B\right)-F_w
    \\
    \gamma = -\left(\alpha k^2 +s +ikF_u\right).
\end{array}
    \label{eq:alphabetagamma}
\end{equation}
%%%%%%%%%%%%%%%%%%%%%%%%%%%%%%%%%%%%%%
The three boundary conditions for Eq.~\eqref{eq:fpafterLFt}are: 

(I) For $w\to\infty$, 
%%%%%%%%%%%%%%%%%%%%%%%%%%%%%%%%%%%%%%
\begin{equation}
\hat{\tilde{P}}(w)\to 0 \qquad (w\to\infty)
    \label{eq:bcnum1}
\end{equation}
%%%%%%%%%%%%%%%%%%%%%%%%%%%%%%%%%%%%%

(II) At $w=0$ Eq.~\eqref{eq:zerofluxGen} yields
%%%%%%%%%%%%%%%%%%%%%%%%%%%%%%%
\begin{equation}
2\alpha \frac{\partial}{\partial w}\hat{\tilde{P}}(w)+\eta \hat{\tilde{P}}(w)=0 \qquad (w\to 0)
    \label{eq:bcnum2}
\end{equation}
%%%%%%%%%%%%%%%%%%%%%%%%%%%%%%%%
where $ \eta=-2F_w+ik(D_A-D_B)$. 

(III) Integration over a small range that includes $w^O$ gives
%%%%%%%%%%%%%%%%%%%%%%%%%%%%%%%%%
\begin{equation}
\alpha \left[\frac{\partial}{\partial w}\hat{\tilde{P}}(w^{O^+})-\frac{\partial}{\partial w}\hat{\tilde{P}}(w^{O^-})\right]=-e^{-iku^O}.
    \label{eq:bcnum3}
\end{equation}
%%%%%%%%%%%%%%%%%%%%%%%%%%%%%%%%%

The solution of Eq.~\eqref{eq:fpafterLFt} is obtained by the usual ansatz $\hat{\tilde{P}}\sim e^{rw}$ which leads to the quadratic charachtersitic equation $\alpha r^2+\beta r+\gamma =0$ with two solutions $r_\pm=\left(-\beta\pm\sqrt{\beta^2-4\alpha\gamma}\right)/2\alpha$. 
Since $\operatorname{Re}(\beta) = -F_w>0$ and $\forall \operatorname{Re}(s)>0$  $|\operatorname{Re}( \sqrt{\beta^2-4\alpha\gamma})|>|F_w|$,  only $r_-$ has a negative real part. Therefore Eq.~\eqref{eq:bcnum1} yields that for any $w\geq w^O$  $\hat{\tilde{P}}(w)=Ce^{r_- w}$. 
In the range $0\leq w \leq w^O$ the solution is $\hat{\tilde{P}}(w) = Ae^{r_- w}+Be^{r_+ w}$. 
Then Eq.~\eqref{eq:bcnum2} yields that $B=-q{A}$ where $q=\frac{2\alpha r_-+\eta}{2\alpha r_+ + \eta}$, and $\hat{\tilde{P}}(w)=A(e^{r_-w}-qe^{r_+ w})$ for $0\leq w \leq w^O$. The continuity at $w=w^O$ leads to $C=A(1-q e^{(r_+-r_-)w^O})$ and 
%%%%%%%%%%%%%%%%%%%%%%%%%%%%%%%%%%%%%%
\begin{equation}
\hat{\tilde{P}}(k,w;s)=
\begin{cases}
Ae^{r_- w}\big(1-q e^{(r_+-r_-) w}\big), & 0\le w\le w^0,\\
Ae^{r_- w}\big(1-q e^{(r_+-r_-) w^0}\big), & w\ge w^0,
\end{cases}
    \label{eq:fpsolutioninfl}
\end{equation}
%%%%%%%%%%%%%%%%%%%%%%%%%%%%%%%%%%%%%%%
where 
%%%%%%%%%%%%%%%%%%%%%%%%%%%%%%%%%%%%%%
\begin{equation}
A=\frac{e^{-iku^0}}{\alpha\,q\,(r_--r_+)\,e^{r_+ w^0}}
    \label{eq:Afinalform}
\end{equation}
%%%%%%%%%%%%%%%%%%%%%%%%%%%%%%%%%%%%%%%
is obtained from Eq.~\eqref{eq:bcnum3}. 
Equation~\eqref{eq:fpsolutioninfl} is the Laplace-Fourier transform of the positional probability density $P(u,w;t)$, and it allows us to find the boundary probability mass $c(t)$. 

The boundary probability mass $c(t)$ is defined in the main text as 
%%%%%%%%%%%%%%%%%%%%%%%%%%%%%%%%%%
\begin{equation}
c(t)=\int_{\partial \Omega} P_B(\tilde{x},t)\,dS_{\tilde{x}}
    \label{eq:ctdefinition}
\end{equation}
%%%%%%%%%%%%%%%%%%%%%%%%%%%%%%%%%%
where $P_B(\tilde{x},t)$ is the PDF at the boundary. 
In terms of the PDF $P(u,w;t)$ the expression for $c(t)$ is $c(t)=\int_{-\infty}^\infty P(u,0;t)\,du$. Since $\tilde{P}(k=0,w;t)=\int_{-\infty}^\infty P(u,w;t)\,du$, from Eq.~\eqref{eq:fpsolutioninfl} we obtain the expression for $c(t)$ in Laplace space, 
%%%%%%%%%%%%%%%%%%%%%%%%%%%%%%%%%%%%
\begin{equation}
\hat{c}(s) = \frac{2e^{-\frac{F_w+\sqrt{F_w^2+4\alpha s}}{2\alpha}w^O}}{F_w+\sqrt{F_w^2+4\alpha s}}
    \label{eq:ctLaplace}
\end{equation}
%%%%%%%%%%%%%%%%%%%%%%%%%%%%%%%%%%%%
and the inverse Laplace transform leads to
%%%%%%%%%%%%%%%%%%%%%%%%%%%%%%%%%%%%
\begin{equation}
    \label{eq:ctexpressiont}
    \begin{array}{l}
    c(t)=
    \\
\frac{1}{\sqrt{\pi \alpha t}}\,
\exp\!\left[-\frac{(w^0-|F_w| t)^2}{4\alpha t}\right]
\;+\;
\frac{|F_w|}{2\alpha}\,
\operatorname{erfc}\!\left(\frac{w^0-|F_w| t}{2\sqrt{\alpha t}}\right).
\end{array}
\end{equation}
%%%%%%%%%%%%%%%%%%%%%%%%%%%%%%%%%%%%
From Eq.~\eqref{eq:ctexpressiont}, the short and long time limits are easily obtained. Namely, 
%%%%%%%%%%%%%%%%%%%%%%%%%%%%%%%%%%%
\begin{equation}
c(t)\to 0 \qquad (t\to0)
    \label{eq:ctshortt}
\end{equation}
%%%%%%%%%%%%%%%%%%%%%%%%%%%%%%%%%%%
and
%%%%%%%%%%%%%%%%%%%%%%%%%%%%%%%%%%%
\begin{equation}
c(t)\to 2\frac{|F_w|}{D_A+D_B} \qquad (t\to\infty).
    \label{eq:ctshortt}
\end{equation}
%%%%%%%%%%%%%%%%%%%%%%%%%%%%%%%%%%%
The results hold for any $F_w<0$, and non-zero initial distance between the particles, i.e., $w^O>0$.

\section*{SECTION B. Numerical simulations}

\setcounter{equation}{0}
\renewcommand{\theequation}{B\arabic{equation}}

We simulate the constrained Langevin dynamics using a discrete-time realization of a Skorokhod reflected process corresponding to Eq.~(10) of the main text. The particle positions $x_A(t)$ and $x_B(t)$ evolve under constant forces $F_A,F_B$ and diffusion coefficients $D_A,D_B$, subject to the hard constraint $x_A \ge x_B$.

Time is discretized with step $\Delta t$. Over each step, unconstrained trial positions $(x_A',x_B')$ are generated using an Euler--Maruyama scheme,
\begin{equation}
\begin{array}{ll}
x_A' &= x_A + F_A\,\Delta t + \sqrt{2D_A\,\Delta t}\,\xi_A, \\
x_B' &= x_B + F_B\,\Delta t + \sqrt{2D_B\,\Delta t}\,\xi_B,
\end{array}
\end{equation}
where $\xi_A,\xi_B$ are independent standard Gaussian random variables with zero mean and unit variance. Introducing the trial relative coordinate
\begin{equation}
u' = x_A' - x_B',
\end{equation}
updates with $u' \ge 0$ satisfy the constraint and are accepted. 
If a trial update violates the constraint ($u'<0$), a Skorokhod correction is applied. 
The correction acts along the  reflection direction
%%%%%%%%%%%%%%%%%%%%%%%%%%%
\begin{equation}
\mathbf{r}(\mathbf{x})
=\frac{\mathbf{D}\,\mathbf{n}(\mathbf{x})}
       {\mathbf{n}(\mathbf{x})^{\!\top}\mathbf{D}\,\mathbf{n}(\mathbf{x})},
\qquad \mathbf{x}\in\partial\Omega ,
\label{eq:rdirection}
\end{equation}
%%%%%%%%%%%%%%%%%%%%%%%%%%%
provided by Eq.~(6) of the main text while $\mathbf{x}=(x_A,x_B)$. 
The diffusion tensor  $\mathbf{D}$ is provided by   Eq.~(12) and $\mathbf{n}=(1,-1)$ is the normal to the boundary $\partial\Omega=\{x_A=x_B\}$.

The scalar Skorokhod increment $\Delta L$ is chosen such that the normal component of the displacement $\mathbf{r}\,\Delta L$ exactly compensates the boundary violation. 
Since the distance of the trial configuration from the boundary in the normal direction is the absolute value of $u'=x_A'-x_B'$, the condition that the corrected configuration lies on the boundary reads
%%%%%%%%%%%%%%%%%%%%%%%
\begin{equation}
u' + \mathbf{n}^{\!\top}\mathbf{r}\,\Delta L = 0 .
\end{equation}
%%%%%%%%%%%%%%%%%%%%%%%%
By construction of $\mathbf r$, one has $\mathbf{n}^{\!\top}\mathbf{r}=1$, and therefore
\begin{equation}
\Delta L = \max(0,-u') .
\end{equation}

The corrected positions are then
%%%%%%%%%%%%%%%%%%%%%%
\begin{equation}
\begin{array}{ll}
x_A &= x_A' + r_A\,\Delta L \\
x_B &= x_B' + r_B\,\Delta L 
\end{array}
\end{equation}
%%%%%%%%%%%%%%%%%%%%%%%%%%%
where, from Eq.~\eqref{eq:rdirection},
%%%%%%%%%%%%%%%%%%%%%%%
\begin{equation}
\begin{array}{l}
r_A=D_A/(D_A+D_B)
\\
r_B=-D_B/(D_A+D_B).
\end{array}{l}
    \label{eq:rArB}
\end{equation}
%%%%%%%%%%%%%%%%%%%%%%%

By construction, $\Delta L\ge0$, acts only when the boundary is crossed, and places the configuration exactly on the boundary. This realizes the minimal Skorokhod reflection, i.e., the smallest nonnegative correction compatible with the constraint. In the limit $\Delta t\to 0$, this scheme converges to the solution of the Fokker–Planck equation with zero-flux boundary conditions used in the analytical treatment.